# Multiplexing spectral line shape of waveguide transmission by photonic spin-orbit interaction

Yuqiong Cheng,[1] Wanyue Xiao,[2] and Shubo Wang[1,2,*]
[1]*Department of Physics, City University of Hong Kong, Tat Chee Avenue, Kowloon, Hong Kong, China*
[2]*City University of Hong Kong Shenzhen Research Institute, Shenzhen, Guangdong 518057, China*

Manipulating the spectral line shape exhibits great potential in realizing active optical circuits with switching, sensing, and modulation capabilities. Exploring unusual line shapes, such as Fano resonance and electromagnetically induced transparency (EIT), has attracted substantial interest. Conventional methods of engineering the spectral line shape have limited tunability and face challenges in multiplexing different spectral line shapes. Here, we propose and numerically demonstrate a new mechanism to tailor the transmission line shape almost at will by exploiting the interference of frequency-dependent chiral dipolar states in two helix particles sitting above a dielectric waveguide. We show that, by tuning the polarization of the chiral dipoles and exploiting transverse spin-orbit interaction, one can control the asymmetric Pancharatnam-Berry geometric phase for the excited guided waves propagating in opposite directions. The interference of the guided waves respectively excited by the two particles can give rise to transmissions with various line shapes, including Lorentzian-like, antiresonance-like, Fano-like, and EIT-like line shapes, which carry an intriguing property of line shape-momentum locking, i.e., the transmissions in opposite directions have different line shapes. Our findings open new possibilities for multiplexed and multifunctional nanophotonic designs with unprecedented capability of spectral-line shaping. The proposed structures can be conveniently integrated with optical circuits for on-chip applications.

## I. INTRODUCTION

Lorentzian resonance, antiresonance, Fano resonance, and electromagnetically induced transparency (EIT) can give rise to different spectral line shapes with intriguing physics and have been extensively studied in microwave and optical systems [1-4]. Tailoring the spectral line shapes promises important applications in designing multifunctional photonic devices for optical sensing, switching, filtering, and modulations [5-9]. Lorentzian resonance exhibits enlarged amplitude at the resonance frequency and serves as the building block for various fascinating optical functionalities and applications [10-12]. Antiresonance exhibits vanished amplitude accompanied by an abrupt negative phase shift. It originates from the destructive interference between coherent drive and interactions in the strong-coupling regime and can be employed to characterize complex integrated quantum circuits [13]. Fano resonance occurs when a discrete state interferes with a continuum state, giving rise to an asymmetric line shape with sharp transition between a dip and a peak [14-16]. It can be applied to realize optical switches, optical sensors, ultrathin circular polarizers, and nonlinear and slow-light devices [17-21]. The EIT arises from the weak coupling between a bright state and a dark state, where the destructive interference between different excitation channels cancels the losses out and results in a narrow transparency window [22]. The highly dispersive nature of EIT paves the way for achieving quality-factor enhancement, photonic information storage, and light stopping [23-27].

Classical analogs of the Fano and EIT spectral line shapes are usually realized by tuning the near-field coupling of neighboring resonators, directly mimicking the quantum interference phenomena in coupled photonic systems [28-30]. The Fano line shape can emerge when high-order modes interact with continuum or dipolar mode in symmetry-breaking systems and hybrid nanostructures [31-33]. It can also appear in photonic crystals with coupling between narrow Bragg resonances and broad Mie or FP resonances, as well as in plasmonic and dielectric metamaterials with collective Fano interference [34, 35]. In these coupled resonator systems, EIT can be achieved as a special case of Fano resonance when the bright and dark modes resonate at the same frequency [36-39]. The transition between Fano and EIT line shapes can be realized by tuning the geometry-dependent resonant states, phase delays, and coupling strength of the resonators (i.e., over-coupling or under-coupling) [40, 41]. In addition, EIT can be modulated by optical chiral states at an exceptional point in an indirectly coupled resonator system [42]. Recently, engineering the spectral line shapes has attracted increasing attention. This can be achieved by introducing additional degrees of freedom, such as the mode symmetry in double-Fano metasurfaces, the backscattering and backcoupling in silicon ring resonators [43, 44], etc. However, these methods have limited tunability and cannot achieve the multiplexing of different spectral line shapes in the same optical structure, which hinders their applications in designing compact and multifunctional optical devices.

Here, we propose a mechanism to realize the channel-dependent multiplexing of spectral line shapes based on photonic spin-orbit interaction (SOI). Photonic SOI refers to the interplay between spin and orbital angular momentum of light [45, 46], which can give rise to numerous intriguing phenomena and properties such as optical spin-Hall effect [47, 48], spin-dependent vortex generation [49], and lateral optical force [50]. Our proposed mechanism relies on the spin-momentum locking property of transverse SOI associated with evanescent waves [51-54]. It is usually employed in the spin-dependent directional excitation of surface/guided waves (i.e., power modulation) [55]. Here, we show that it can be applied to achieve polarization-dependent asymmetric Pancharatnam-Berry (PB) phase for the guided waves, which serves as the key mechanism for the versatile channel-dependent spectral-line shaping. Our system consists of two chiral particles sitting above a dielectric waveguide. By selectively exciting the dipole modes in the particles and controlling their polarizations, one can achieve a transition of the excited guided wave from negative phase shift to positive phase shift at the resonance frequency. The interference between the two guided waves excited respectively by the two particles can give rise to waveguide transmissions with Lorentzian-like, antiresonance-like, Fano-like, and EIT-like line shapes. In addition, the transmissions in opposite guided wave channels have entirely different line shapes, e.g., Lorentzian-like in one direction and Fano-like in the opposite direction. This enables the switching of different line shapes in different transmission channels of the waveguide without changing the resonator geometry or working frequencies, which paves the way to realizing channel-dependent multifunctionalities for optical communications.

---

*shubwang@cityu.edu.hk

The paper is organized as follows. In Sec. II, we propose and explain the mechanism for the phase modulation by the synergy of transverse SOI and PB geometric phase. In Sec. III, we implement the mechanism in the helix-waveguide coupling system and demonstrate the polarization- and channel-dependent phase shift of the excited guided wave. In Sec. IV, we realize the versatile channel-multiplexed manipulation of transmission line shapes by engineering the interference of guided waves. We draw the conclusion in Sec. V.

## II. MECHANISM

The spectral line shape is attributed to the interferences of all excited wave channels. Fano resonance corresponds to the interference of a resonant channel and continuum, where the former undergoes a phase change of $\pi$ at the resonance frequency while the phase of the later remains unchanged, resulting in a sharp transition between the constructive interference (peak) and the destructive interference (dip). Similarly, EIT corresponds to the interference of two resonant channels with opposite phase shifts and is featured by a narrow transparent window due to the destructive interference [28]. Thus, the independent modulation of the phase shifts in the wave channels is critical to realizing different spectral line shapes. We will show that this can be achieved by the synergy of the PB geometric phase and transverse SOI.

To illustrate the mechanism, we consider a two-dimensional (2D) electric dipole with two Cartesian components $\mathbf{p}_{2D} = (p_x, p_z)$ sitting above a dielectric slab waveguide, as shown in Fig. 1(a). Here, we assume that $p_x$ and $p_z$ both have a Lorentzian frequency dependence with different resonance frequencies $\omega_1$ and $\omega_2$ as well as damping $\gamma_1$ and $\gamma_2$: [56]

$$p_x = \frac{p_{x0}}{\omega_1^2 - \omega^2 - 2i\gamma_1\omega}, \quad (1)$$

and

$$p_z = \frac{p_{z0}}{\omega_2^2 - \omega^2 - 2i\gamma_2\omega}, \quad (2)$$

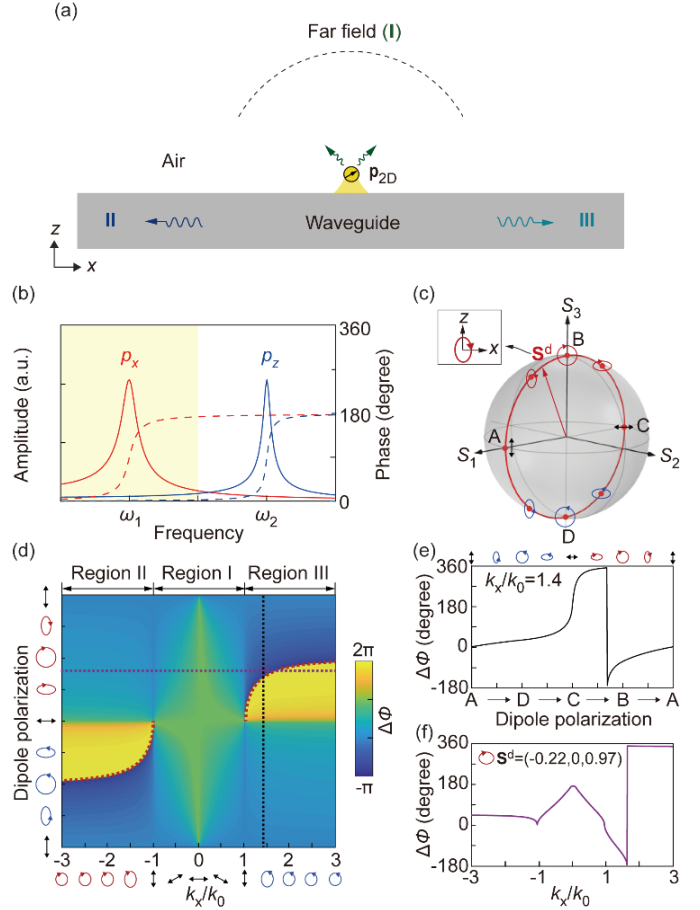

FIG. 1. Phase shift modulation enabled by transverse SOI. (a) The electric dipole $\mathbf{p}_{2D}$ can couple to three different wave channels: far-field radiation channel (I), and guided wave channels (II and III). (b) The amplitudes and phases of two dipole components $p_x$ and $p_z$ with different resonance frequencies. (c) Evolution of dipole polarization on the Poincaré sphere when the ratio of dipole components changes. The inset shows the corresponding dipole polarization ellipse. The blue and red colors of the polarization ellipses denote opposite spin directions. (d) The phase shift $\Delta\phi$ as a function of the dipole polarization and the wavevector $k_x/k_0$ of the wave channel. The black dashed line corresponds to $k_x/k_0 = 1.4$, and purple dashed line corresponds to $\mathbf{S}^d = (-0.22, 0, 0.97)$. The polarization of the wave channel is shown by the polarization ellipses below the horizontal axis. (e) The phase shift $\Delta\phi$ as a function of dipole polarization when $k_x/k_0 = 1.4$, corresponding to the black dashed line in (d). (f) The phase shift $\Delta\phi$ as a function of $k_x/k_0$ when $\mathbf{S}^d(\omega_1) = (-0.22, 0, 0.97)$, corresponding to the purple dashed line in (d).

where $p_{x0}$ and $p_{z0}$ are the amplitudes determined by the external excitation. Figure 1(b) shows the spectra of the two dipole components for $\omega_1 < \omega_2$. Near the resonance frequency of $p_x$ (in the yellow region), the amplitude of $p_x$ (denoted by the red solid line) has a Lorentzian line shape and the phase of $p_x$ (denoted by the red dashed line) experiences 180-degree shift, while the amplitude and phase of $p_z$ (denoted by the blue solid and dashed lines, respectively) are approximately unchanged. Thus, the phase difference between $p_x$ and $p_z$ changes from 0 degree to 180 degrees, i.e., the polarization of the dipole $\mathbf{p}_{2D}$ varies with the frequency. Specially, at $\omega = \omega_1$, $p_x$ and $p_z$ have a phase difference of 90 degrees because $p_x$ is on resonance while $p_z$ is off resonance, i.e., the dipole $\mathbf{p}_{2D}$ is elliptically polarized. The polarization of $\mathbf{p}_{2D}$ can be characterized by the Stokes vector $\mathbf{S}^d = (S_1^d, S_2^d, S_3^d)$ with $S_1^d = \frac{|p_z|^2 - |p_x|^2}{|p_z|^2 + |p_x|^2}$, $S_2^d = \frac{2\text{Re}(p_z p_x^*)}{|p_z|^2 + |p_x|^2}$, and $S_3^d = \frac{-2\text{Im}(p_z p_x^*)}{|p_z|^2 + |p_x|^2}$, as denoted by the red arrow in the Poincaré sphere in Fig. 1(c), where the inset shows the corresponding polarization ellipse. The polarization ellipticity of $\mathbf{p}_{2D}$ is decided by the ratio $p_{z0}/p_{x0}$. When $p_{z0}/p_{x0}$ changes, the tail of the Stokes vector $\mathbf{S}^d$ traces out a trajectory on the Poincaré sphere, denoted by the red circle in the $S_1$-$S_3$ plane, where the polarization ellipses of some representative points are shown. Particularly, the points A, B, C, and D denote the z-polarized linear dipole, right-handed circular dipole, x-polarized linear dipole, and left-handed circular dipole, respectively.

The fields of the dipole $\mathbf{p}_{2D}$ can couple to the far-field radiation channels in the region I, the left-propagating guided wave channel in the region II, and the right-propagating guided wave channel in the region III, as shown in Fig. 1(a). The coupling strongly depends on the polarization match between the dipole and the wave channels. The corresponding coupling coefficient given by the coupled mode theory (CMT) is $\kappa(\omega) \propto \mathbf{p}_{2D} \cdot \overline{\mathbf{E}}^* \propto (-p_x k_z^* + p_z k_x^*)$ [52], where $\overline{\mathbf{E}}$ denotes the eigen electric field of the wave channels with the form $\overline{\mathbf{E}} = (-k_z \hat{\mathbf{e}}_x + k_x \hat{\mathbf{e}}_z)e^{(ik_x x + ik_z z)}$, $k_x$ and $k_z$ are wavevector components in x and z directions satisfying $k_x^2 + k_z^2 = k_0^2 = \omega^2/c^2$ and we have neglected the time-harmonic factor $e^{-i\omega t}$. For the guided wave channels, this $\overline{\mathbf{E}}$ field only describes the evanescent tail responsible for the coupling. The polarization of the electric field can be described by the Stokes vector $\mathbf{S}^c = \left(\frac{|k_x|^2 - |k_z|^2}{|k_0|^2}, \frac{-2k_x k_z}{|k_0|^2}, 0\right)$ for the radiation channels and $\mathbf{S}^c = \left(\frac{|k_0|^2}{|k_x|^2 + |k_z|^2}, 0, \frac{-2k_x \text{Im}(k_z)}{|k_x|^2 + |k_z|^2}\right)$ for the guided wave channels propagating in $\pm x$ direction. We note that the guided wave channels are elliptically polarized with the sign of $S_3^c$ (i.e., spin direction) locked to propagating wavevector $k_x$, corresponding to the spin-momentum locking of transverse SOI [55]. Near the resonance frequency $\omega_1$, the coupling coefficient can be approximated as

$$\kappa(\omega) \propto \left( -\frac{p_{x0} k_z^*}{\omega_1^2 - \omega^2 - 2i\gamma_1 \omega} + \frac{p_{z0} k_x^*}{\omega_2^2 - \omega^2} \right), \quad (3)$$

where we have neglected the loss of $p_z$ under the condition $|\omega_2^2 - \omega^2| \gg 2\gamma_2 \omega$. The coupling coefficient $\kappa(\omega)$ directly decides the amplitude and phase of the excited waves. Equation (3) indicates that $\kappa(\omega)$ undergoes a dramatic phase shift when the excitation frequency increases from $\omega < \omega_1$ to $\omega > \omega_1$, and the maximum phase shift depends on the polarization of the dipole $\mathbf{p}_{2D}$ and the wavevector component $k_x$. Figure 1(d) shows the maximum phase shift $\Delta \phi$ for $\omega \in [\omega_1/2, 3\omega_1/2]$ as a function of the dipole polarization and $k_x/k_0$. For the radiation channels in the region I ($|k_x| \leq k_0$) without transverse spin, the phase shift $\Delta \phi$ is symmetric with respect to the dipole spin $S_3^d(\omega_1)$ and $k_x$, with a maximum of ~180 degrees. Increasing $S_1^d(\omega_1)$, corresponding to the ellipse path C → B → A (or C → D → A) on the Poincaré sphere, and increasing $|k_x|$ will both reduce the relative contribution of $p_x$ over $p_z$ and the phase shift of the excited propagating waves. In contrast, for the guided wave channels in the region II ($k_x < -k_0$) and region III ($k_x > k_0$) with transverse spin, $\Delta \phi$ is asymmetric with respect to $S_3^d(\omega_1)$ and $k_x$. Importantly, controlling the dipole polarization or channel wavevector can give rise to almost arbitrary phase shift of the excited guided wave. Specifically, when the dipole and the guided wave channel have opposite spin directions (i.e., opposite polarization handedness denoted by different colors of the polarization ellipses), the maximum phase shift can take the values $\Delta \phi \in [180, 360]$ and $[-180, 0]$. In this case, a jump of $\Delta \phi$ from 360 degrees to -180 degrees appears at $\mathbf{S}^d(\omega_1) = -\mathbf{S}^c$ (i.e., orthogonal polarizations), as marked by the red dashed line, corresponding to the vanished excitation of the guided wave. Differently, when the dipole and guided wave channel have the same spin direction (i.e., the same polarization handedness denoted by the same color of the polarization ellipses), the phase shift can take the values $\Delta \phi \in [0, 180]$. To further understand the dependence of the phase shift on the dipole polarization and wavevector component $k_x$, we show the phase shift $\Delta \phi$ for the cases $k_x/k_0 = 1.4$ and $\mathbf{S}^d(\omega_1) = (-0.22, 0, 0.97)$ in Fig. 1(e) and (f), respectively, which correspond to the black and purple dashed lines in Fig. 1(d). As shown in Fig. 1(e), for the guided wave channels ($|k_x| > k_0$), arbitrary phase shift in the range [-180, 360] can be achieved by varying the dipole polarization along the path A → D → C → B → A. As shown in Fig. 1(f), the phase shift $\Delta \phi$ is asymmetric with respect to opposite $k_x$ for the guided wave channels, i.e., the excited guided waves propagating in $\pm x$ directions have different phase shifts. In Fig. 4 and Fig. 5, we will show that this polarization- and channel-dependent phase shift can be used to flexibly manipulate the transmission line shape.

The underlying mechanism of the phase shift modulation is attributed to the PB geometric phases originating from asymmetric polarization evolutions under transverse SOI. The phase of the coupling coefficient $\kappa(\omega)$ can be attributed to the change of polarization when the dipole couples to the guided wave channels. We illustrate this point for two example cases with dipole polarizations $\mathbf{S}^d(\omega_1) = (-0.6, 0, 0.8)$ and $\mathbf{S}^d(\omega_1) = (-0.22, 0, 0.97)$, which give rise to the positive and negative phase shifts of the guided wave with $k_x > 0$, respectively. When the frequency $\omega$ increases from $\omega_1/2$ to $3\omega_1/2$, the polarization of the dipole with $\mathbf{S}^d(\omega_1) = (-0.6, 0, 0.8)$ traces out a trajectory on the Poincaré sphere, which is denoted by the red curve marked by i → ix in Fig. 2(a), with the point v denoting the dipole polarization at $\omega = \omega_1$. The yellow point marked by $c_\pm$ denotes the polarization of the evanescent tail of the guided wave channel with $\mathbf{S}^c = (0.34, 0, \mp 0.94)$, where the upper (lower) sign corresponds to $k_x/k_0 = 1.4$ ($k_x/k_0 = -1.4$). Choosing a reference point m (0, 0, 1) on the Poincaré sphere, we can determine the geometric phase associated with the polarization evolution from the dipole polarization "i" to the wave channel polarization "$c_{\pm}$", which is given by the solid angle subtended by the area enclosed by the contour "i → $c_{\pm}$ → m → i" [57, 58]. The geometric phase for other dipole polarizations (i.e., ii to ix) can be determined similarly. We note that the variation of the geometric phase (indicated by the evolution of the blue area) is different for $k_x > 0$ (left panel in Fig. 2(a)) and $k_x < 0$ (right panel in Fig. 2(a)) due to the different spin $S_3^c$ of opposite propagating guided waves, which is protected by the spin-momentum locking of transverse SOI. This explains the asymmetric phase shift with respect to $k_x$ in Fig. 1(d). To verify the above explanation, we compare the obtained PB geometric phase with the phase of $\kappa(\omega)$ given by CMT in Eq. (3), as shown in Fig. 2(b). We notice the good agreements for both $k_x > 0$ and $k_x < 0$, demonstrating the validity of our interpretation based on PB geometric phase. Similarly, we verify the geometric phase interpretation for the dipole with polarization $\mathbf{S}^d(\omega_1) =$ (-0.22, 0, 0.97) in Figs. 2(c) and (d), which again exhibits good consistency with the CMT results. Compared with Fig. 2(a), it is clear that different dipole polarization will give rise to different evolution of the geometric phase (i.e., blue area). The PB phase provides a clear geometric interpretation for the dependence of the phase shift on the dipole polarization and the guided wave channel.

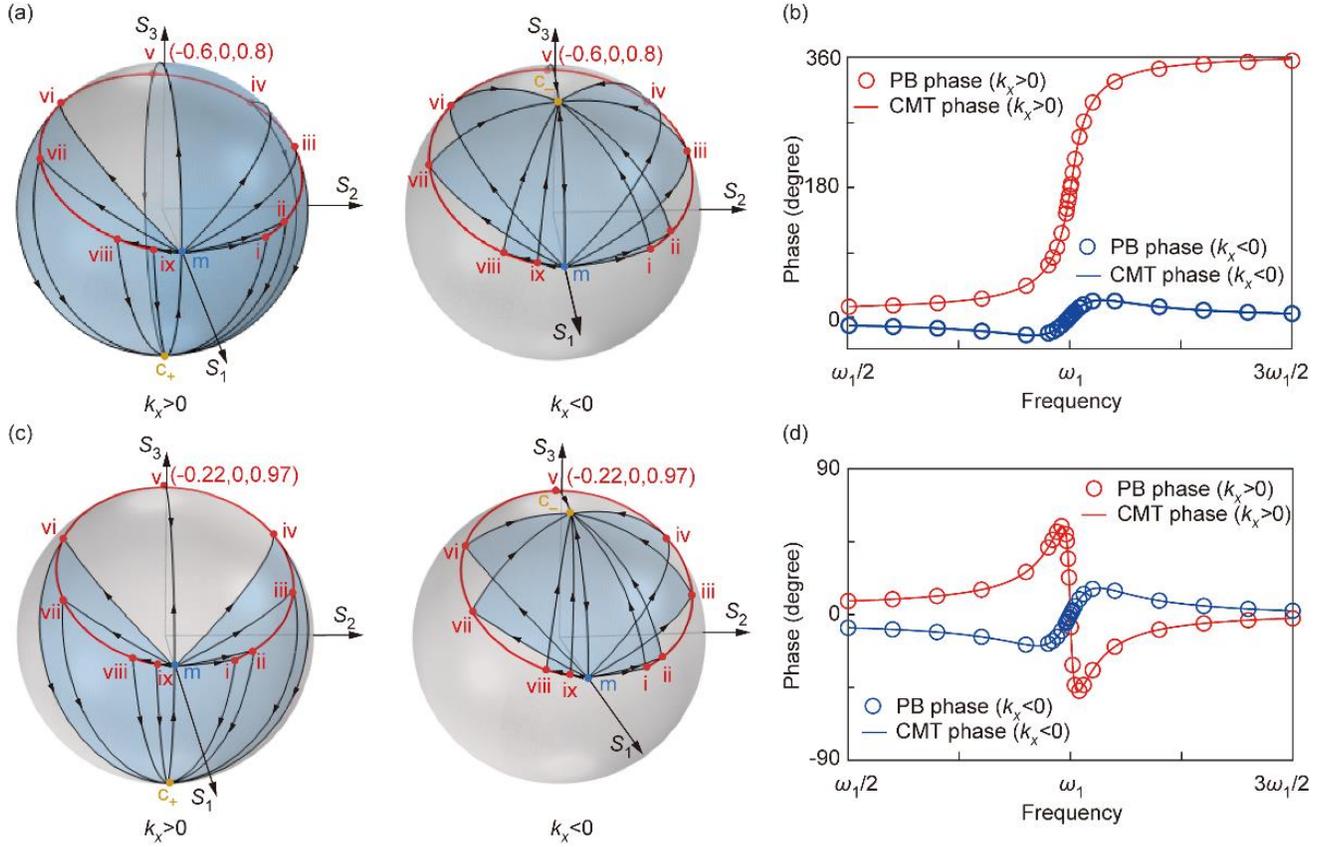

FIG. 2. A geometric phase interpretation of the phase shift. (a) Polarization evolution on the Poincaré sphere for the electric dipole with Stokes vector $\mathbf{S}^d(\omega_1) =$ (-0.6, 0, 0.8) coupled to the waveguide. The red curves marked by i → ix denote the polarization variation of the electric dipole with increasing frequency. The point $c_{\pm}$ denotes the polarization of the evanescent tail of the guided wave channels with wavevector component $k_x/k_0 = \pm 1.4$. The blue areas enclosed by the black arrow contour give the PB geometric phase for the guided wave excited by different dipole polarizations at different frequencies. (b) Comparison between the PB phase and the CMT phase for the electric dipole with $\mathbf{S}^d(\omega_1) =$ (-0.6, 0, 0.8). (c) Polarization evolution on the Poincaré sphere for the electric dipole with Stokes vector $\mathbf{S}^d(\omega_1) =$ (-0.22, 0, 0.97) coupled to the waveguide. (d) Comparison between the PB phase and the CMT phase for the electric dipole with $\mathbf{S}^d(\omega_1) =$ (-0.22, 0, 0.97).

### III. REALIZATION OF PHASE MODULATION

To realize the above mechanism of phase modulation, we consider a metal helix sitting over a silicon ($\varepsilon_{si} = 12$) waveguide under the excitation of an electromagnetic plane wave, as shown in Fig. 3(a). The helix is right-handed with two turns and has pitch $P = 75$ nm, outer radius $R = 46$ nm, and inner radius $r = 11$ nm. It is made of silver with relative permittivity characterized by the Drude model $\varepsilon_{Ag} = 3.92 - \omega_p^2/(\omega^2 + i\omega\omega_t)$, where $\omega_p = 1.33 \times 10^{16}$ rad/s and $\omega_t = 2.73 \times 10^{13}$ rad/s [59]. The waveguide has a rectangular cross section of $w \times t = 630$ nm $\times$ 315 nm.

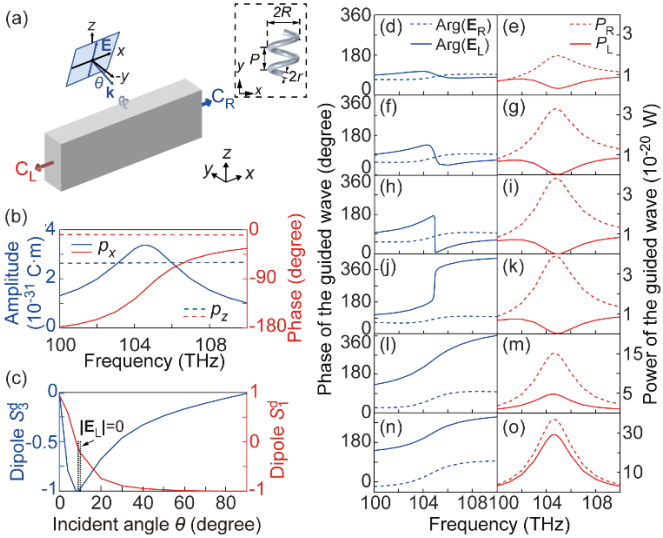

FIG. 3. Phase manipulation of the guided wave excited by a helix. (a) Helix-waveguide coupling configuration. $C_L$ and $C_R$ respectively represent the left and right guided wave channels. (b) The amplitudes and phases of the dipoles $p_x$ and $p_z$ induced in the helix as a function of frequency. (c) Stokes parameters of the electric dipole induced in the helix as a function of the incident angle $\theta$ at the frequency 104.9 THz. The evolution of phase (left column) and power (right column) of the guided wave for (d, e) $\theta = 4$ degrees, (f, g) $\theta = 8$ degrees, (h, i) $\theta = 9.4$ degrees, (j, k) $\theta = 9.6$ degrees, (l, m) $\theta = 30$ degrees, and (n, o) $\theta = 70$ degrees.

The distance between the center of the helix and the waveguide's upper surface is 125 nm. The incident plane wave is linearly polarized with the electric field $\mathbf{E}_{inc} = (-\sin\theta\hat{y} + \cos\theta\hat{z})E_0 e^{-ik_0(\cos\theta y + \sin\theta z)}$, where $\theta$ is the incident angle between the wavevector $\mathbf{k}$ and $-y$ direction on the $zoy$ plane, and the time-harmonic factor $e^{-i\omega t}$ is neglected. For simplicity, we assume the plane wave only incidents on the helix and the background scattering by the waveguide is neglected. All the numerical simulations in the following are conducted by using finite-element package COMSOL.

Under the excitation of the plane wave, the charges in the helix give rise to the electric dipoles $p_x$ and $p_z$ that dominate the coupling with the waveguide [60]. Figure 3(b) shows the induced dipoles for $\theta = 10$ degrees obtained via numerical simulations. We notice that $p_x$ and $p_z$ originate from two different resonant modes with different resonance frequencies. The dipole component $p_x$ resonates at 104.6 THz accompanied by a phase shift of $\pi$, while the non-resonant $p_z$ is approximately a constant, corresponding to the scenario in Fig. 1(b). At the resonance frequency of $p_x$, we have $\mathrm{Arg}(p_z) - \mathrm{Arg}(p_x) = \pi/2$, and $|p_z|/|p_x|$ can be tuned by varying the incident angle $\theta$ [60]. Figure 3(c) shows the polarization of the induced electric dipole $\mathbf{p} = (p_x, p_z)$ at the frequency $\omega_1/2\pi = 104.9$ THz for different incident angles. When $\theta$ increases, the dipole spin $S_3^d(\omega_1)$ (denoted by the blue line) first decreases and then increases, and $S_1^d(\omega_1)$ monotonically decreases, which approximately corresponds to the dipole polarization evolution "A → D → C" in Fig. 1(c). In the considered frequency range, the dipole will excite the fundamental guided waves propagating in both $-x$ and $+x$ directions, corresponding to the left and right guided wave channels ($C_L$ and $C_R$). At about $\theta = 9.5$, the dipole polarization is orthogonal to the polarization of the evanescent tail of $C_L$, leading to perfect unidirectional coupling to $C_R$. We denote the electric field of the left- (right-) propagating guided wave as $\mathbf{E}_L$ ($\mathbf{E}_R$), and compute the phase (i.e., $\mathrm{Arg}(\mathbf{E}_L)$ and $\mathrm{Arg}(\mathbf{E}_R)$) and power (i.e., $P_L$ and $P_R$) of the fields. Figure 3(d-o) shows the phase (blue lines) and power (red lines) under different incident angle $\theta$. We find that a higher power directionality results in a larger difference between the phase shifts of the guided waves in opposite guided wave channels. For $\theta = 0$ degree or $\theta = 90$ degrees, the electric dipole $\mathbf{p}$ is linearly polarized and symmetrically excites the guided waves. For $0 < \theta < 90$ degrees, the electric dipole $\mathbf{p}$ is elliptically polarized and results in asymmetric excitation of the guided waves. As shown by the blue dashed lines, with the increase of $\theta$, the phase shift of the right-propagating guided wave increases and approaches 180 degrees. The power $P_R$ denoted by red dashed lines follows the Lorentzian-like line shape with its peak rising with $\theta$. This is attributed to the increase of $p_x$, the contribution of which dominates in the excited guide wave at large values of $\theta$. Differently, when the incident angle $\theta$ increases, the phase shift of the left-propagating guided wave (denoted by the blue solid lines) changes from negative to positive. The transition point appears between $\theta = 9.4$ degrees and $\theta = 9.6$ degrees, corresponding to Fig. 3(h) and 3(j), respectively, where near-perfect unidirectional coupling happens. At $\theta = 9.4$ degrees, the phase of the guided wave $\mathrm{Arg}(\mathbf{E}_L)$ undergoes a negative phase shift of 180 degrees near the resonance frequency, while $\mathrm{Arg}(\mathbf{E}_L)$ at $\theta = 9.6$ degrees experiences a positive phase shift of 360 degrees. Besides, the line shape of the power $P_L$ experiences a transition from antiresonance-like to Lorentzian-like, as shown by the red solid lines. These results agree well with the analytical results in Fig. 1. Thus, the phase shift in the guided wave channels can be flexibly controlled by tuning the polarization state of the helix, which serves as the key mechanism for manipulating the transmission line shapes.

## IV. MANIPULATION OF SPECTRAL LINE SHAPE

We apply two helix particles ($s_1$ and $s_2$) coupled to the waveguide to realize different spectral line shapes, as shown in Fig. 4(a). The inset shows the four wave channels: light couples from the helix $s_1$ to the left and right guided wave channels ($E_1^L$ and $E_1^R$), and light couples from the helix $s_2$ to the left and right guided wave channels ($E_2^L$ and $E_2^R$). The phase shifts of the $E_1^{L,R}$ and $E_2^{L,R}$ are respectively determined by the dipole polarization of the helix particles $s_1$ and $s_2$. The power in the four wave channels ($P_{1,2}^{L,R}$) can be controlled by the particle-waveguide separation distance $\Delta h_1$ and $\Delta h_2$. Through tuning the dipole polarizations of the helix particles and the particles' position, we can control the interference between $\mathbf{E}_1^L$ and $\mathbf{E}_2^L$ ($\mathbf{E}_1^R$ and $\mathbf{E}_2^R$) in the left (right) guided wave channel and manipulate the transmission line shape. It is important to note that this mechanism is fundamentally different from the conventional methods employing the near-field coupling of neighboring resonators. Here, the line-shape tunability is attributed to the interference of helix-waveguide coupling channels with engineered phase and power relations.

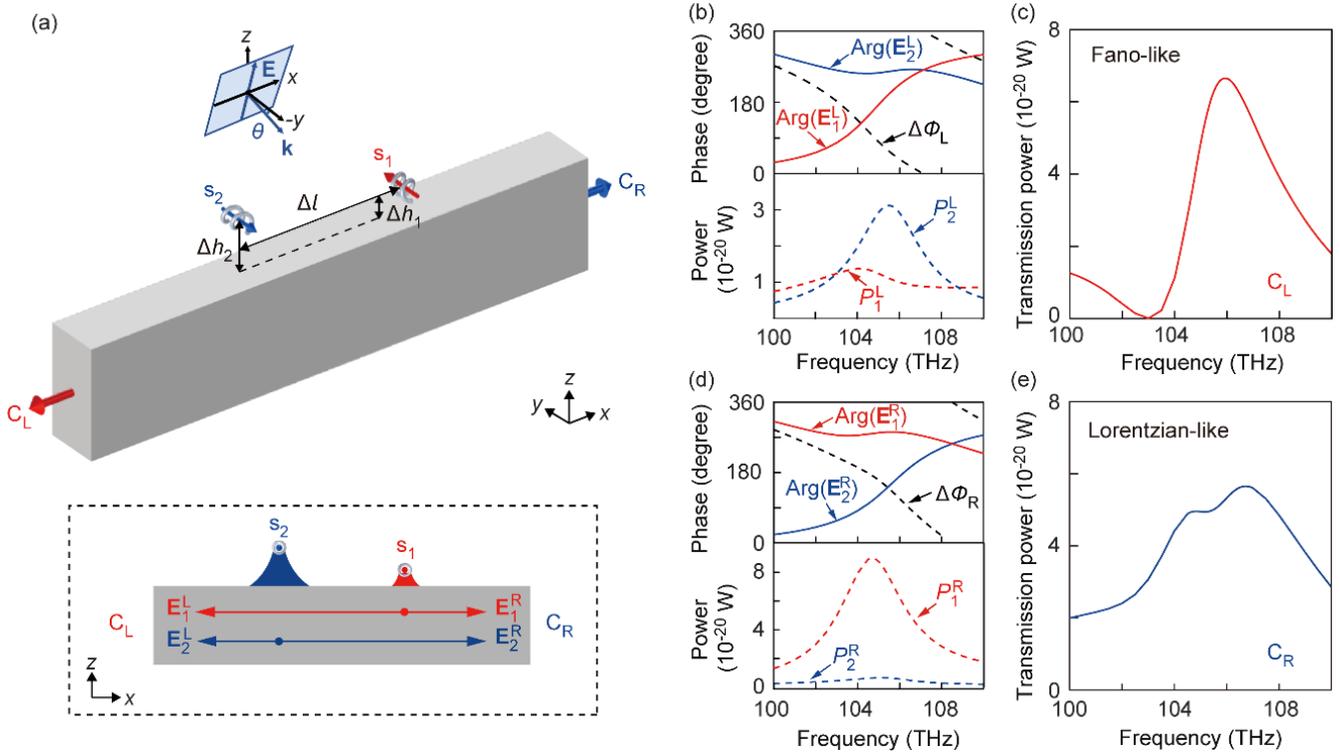

FIG. 4. Fano-like and Lorentzian-like line shapes realized by the bi-helix-waveguide coupling system. (a) Two helix particles $s_1$ and $s_2$ with opposite handedness are placed over the waveguide. The $\Delta h_1$ ($\Delta h_2$) denotes the separation distance between $s_1$ ($s_2$) and the waveguide's upper surface, and $\Delta l$ denotes the separation distance between $s_1$ and $s_2$ along $x$ direction. The inset shows the four wave channels. (b) The phase and power of the guided waves $\mathbf{E}_1^L$ and $\mathbf{E}_2^L$. (c) Fano-like transmission line shape in the left guided wave channel $C_L$. (d) The phase and power of the guide waves $\mathbf{E}_1^R$ and $\mathbf{E}_2^R$. (e) Lorentzian-like transmission line shape in the right guided wave channel $C_R$.

Figure 4(a) shows the system with two helix particles of opposite handedness—$s_1$ is right-handed and $s_2$ is left-handed, which can simultaneously give rise to Fano-like and Lorentzian-like line shapes in opposite guided wave channels. The separation between $s_1$ ($s_2$) and waveguide's upper surface is $\Delta h_1 = 125$ nm ($\Delta h_2 = 302$ nm), ensuring that the excited wave channels $\mathbf{E}_1^L$ and $\mathbf{E}_2^L$ have comparable powers for interference. The separation between $s_1$ and $s_2$ along $x$ direction is $\Delta l = 2400$ nm. Under the linearly polarized incidence with $\theta = 20$ degrees, the two helix particles $s_1$ and $s_2$ can generate opposite dipole spins and induce opposite asymmetric coupling with the waveguide, giving rise to channel-dependent phase shift and power. Figure 4(b) shows the phase and power of the excited guided waves $\mathbf{E}_1^L$ and $\mathbf{E}_2^L$ as a function of frequency. The phase $\text{Arg}(\mathbf{E}_1^L)$ (denoted by the red line) experiences a shift of nearly 180 degrees within the frequency range [103 THz, 107 THz], while the phase $\text{Arg}(\mathbf{E}_2^L)$ (denoted by the blue line) remains approximately unchanged in this frequency range. The phase difference $\Delta\phi_L = \text{Arg}(\mathbf{E}_1^L) - \text{Arg}(\mathbf{E}_2^L)$ (denoted by the black dashed line) reaches 180 degrees at 103 THz and 0 degree at 107 THz. Meanwhile, the power of the two channels, i.e., $P_1^L$ and $P_2^L$, intersect at around 103 THz and 108 THz. Thus, the interference of $\mathbf{E}_1^L$ and $\mathbf{E}_2^L$ switches sharply from destructive to constructive, giving rise to a Fano-like line shape in the left guided wave channel, as shown in Fig. 4(c). We note that the sharpness of the Fano asymmetric line shape is usually limited by the non-radiative (Ohmic) losses in conventional systems [34]. Our proposed mechanism can break this limitation and increase the sharpness through engineering the polarization match between the dipole and guided wave channel. Figure 4(d) shows the phase and power of the channels $\mathbf{E}_1^R$ and $\mathbf{E}_2^R$. Interestingly, the phase $\text{Arg}(\mathbf{E}_1^R)$ (denoted by the red line) remains approximately unchanged in the frequency range [103 THz, 107 THz], while the phase $\text{Arg}(\mathbf{E}_2^R)$ (denoted by the blue line) undergoes a positive shift of nearly 180 degrees. Since the power $P_1^R$ is much larger than $P_2^R$, the interference between $\mathbf{E}_1^R$ and $\mathbf{E}_2^R$ is dominated by the contribution of the helix $s_1$, giving rise to a Lorentzian-like line shape in the right guided wave channel, as shown in Fig. 4(e).

Figure 5 illustrates the configuration for realizing both EIT-like and antiresonance-like transmission line shapes,

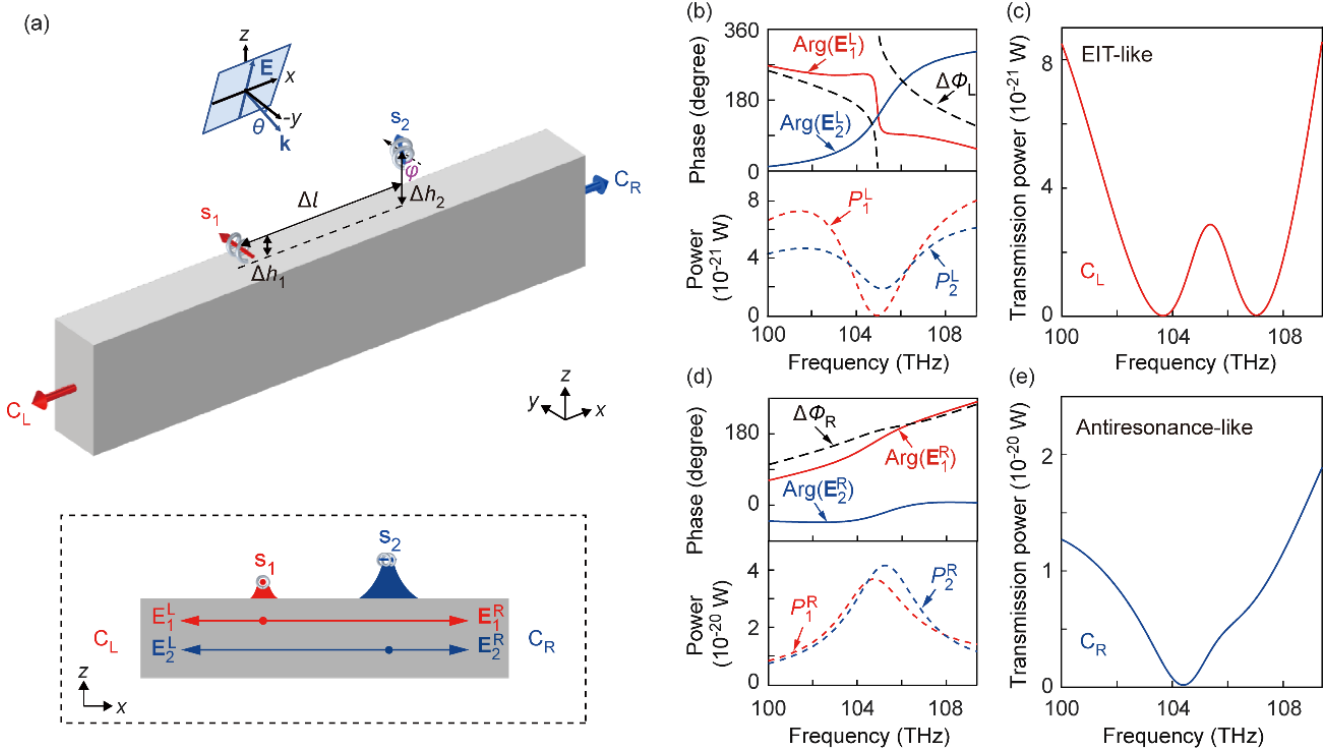

FIG. 5. EIT-like and antiresonance-like line shapes realized by the bi-helix-waveguide coupling system. (a) Two right-handed helix particles $s_1$ and $s_2$ with different axis orientations are placed over the waveguide. The $\Delta h_1$ ($\Delta h_2$) denotes the separation distance between $s_1$ ($s_2$) and the waveguide's upper surface, and $\Delta l$ denotes the separation distance between $s_1$ and $s_2$ along $x$ direction. The center axis of $s_2$ is rotated by angle $\varphi$ with respect to $y$-axis. The inset shows the four wave channels. (b) The phase and power of the excited guided waves $\mathbf{E}_1^L$ and $\mathbf{E}_2^L$. (c) EIT-like transmission line shape in the left guided wave channel $C_L$. (d) The phase and power of the guide waves $\mathbf{E}_1^R$ and $\mathbf{E}_2^R$. (e) Antiresonance-like transmission line shape in the right guided wave channel $C_R$.

where we choose two right-handed helices with different axis orientations. The helix $s_1$ is placed above the waveguide's upper surface with $\Delta h_1 = 125$ nm and its center axis (denoted by the red arrow) is parallel to $y$-axis. The helix $s_2$ is placed above the waveguide's upper surface with $\Delta h_2 = 175$ nm and its center axis (denoted by the blue arrow) is rotated by the angle $\varphi = 8$ degrees with respect to $y$-axis on the $xy$-plane. Here, the rotation of the helix $s_2$ contributes to the adjustment of its dipole polarization. Their separation distance along $x$ direction is $\Delta l = 3065$ nm. Figure 5(b) shows the phase and power of the excited guide waves $\mathbf{E}_1^L$ and $\mathbf{E}_2^L$ under linearly polarized incidence with $\theta = 9$ degrees. The phase $\mathrm{Arg}(\mathbf{E}_1^L)$ (denoted by the red line) experiences a negative shift of nearly 180 degrees when the frequency increases from 103.6 THz to 107 THz, while $\mathrm{Arg}(\mathbf{E}_2^L)$ (denoted by the blue line) undergoes a positive shift of nearly 180 degrees. The phase difference $\Delta \phi_L = \mathrm{Arg}(\mathbf{E}_1^L) - \mathrm{Arg}(\mathbf{E}_2^L)$ (denoted by the black dashed line) evolves from 180 degrees at 103.6 THz, and 0 degree at 105 THz, to 180 degrees at 107 THz. Meanwhile, the power $P_1^L$ intersects with $P_2^L$ at 103.6 THz and 107 THz, generating perfect destructive interferences at the two frequencies. This gives rise to an EIT-like line shape in the left guided wave channel featured by a peak between two dips, as shown in Fig. 5(c). The quality factor of the transparency resonance is about 70, which is much higher than that of the dipole resonance of the helix (~15) shown in Fig. 3(b). Figure 5(d) shows the phase and power of the guided waves $\mathbf{E}_1^R$ and $\mathbf{E}_2^R$. The phase difference of $\mathbf{E}_1^R$ and $\mathbf{E}_2^R$ (i.e., $\Delta \phi_R$, denoted by the black dashed line) mainly originates from the dispersion of the waveguide mode, and it reaches 180 degrees at around 104 THz. Moreover, the power $P_1^R$ and $P_2^R$ almost coincide with each other. The interference of $\mathbf{E}_1^R$ and $\mathbf{E}_2^R$ gives rise to an antiresonance-like transmission line shape in the right guided wave channel, as shown in Fig. 5(e), which exhibits nearly zero power at 104.2 THz.

## V. CONCLUSION

In conclusion, we analytically and numerically demonstrate the channel-dependent manipulation of the transmission spectral line shape in a bi-helix-waveguide coupling system. By selectively exciting the orthogonal dipole modes of the helix and tuning its polarization, we achieve almost arbitrary phase shift of the excited guided waves based on PB geometric phase and transverse SOI. We apply the interference of the excited guided waves with independent phase modulations to tailor the transmission line shape in the waveguide. This gives rise to the multiplexing of channel-dependent spectral line shapes and a remarkable property of line shape-momentum locking, i.e., the waveguide transmission in opposite channels have different spectral line shapes. The proposed mechanism enables significant flexibility of engineering the spectral line shapes for a plethora of applications in photonic integrated circuits, quantum information processing, and nonlinear optics.

**ACKNOWLEDGMENTS**

The work described in this paper was supported by the National Natural Science Foundation of China (No. 11904306) and the Research Grants Council of the Hong Kong Special Administrative Region, China (Nos. C6013-18G, CityU 11301820, and CityU 11306019).